# Quantitative Theory of Meaning. Application to Financial Markets. EUR/USD case study.

Inga Ivanova[1], Grzegorz Rzadkowski[2] and Loet Leydesdorff[3]

**Abstract**

**Purpose** – The paper focuses on the link between information, investors' expectations and market price movement. EUR/USD market is examined from communication-theoretical perspectives on the dynamics of information and meaning.
**Design/methodology/approach** –We build upon the quantitative theory of meaning as a complement to the quantitative theory of information. Different groups of investors entertain different criteria to process information, so that the same information can be supplied with different meanings. Meanings shape investors' expectations which are revealed in market asset price movement. This dynamics can be captured by non-linear evolutionary equation. We use a computationally efficient technique of logistic Continuous Wavelet Transform (CWT) to analyze EUR/USD market.
**Findings** – The results reveal the latent EUR/USD trend structure which coincides with the model predicted time series indicating that proposed model can adequately describe some patterns of investors' behavior.
**Originality/value** – To our best knowledge this is the first paper where investors' expectations and market assets price movement are analyzed on the base of quantitative theory of meaning.
**Practical implication** – Proposed methodology can be used to better understand and forecast future market assets' price movement.
**Social implication** – Information is a universal concept. From this viewpoint communication of information in financial markets doesn't much differ from communication of information in other social systems. The same conceptual tools can be applied to study other complex social systems with similar topology.

**Paper type** Research paper
**JEL Classification** C22, G10, G15, G40, E32, D01, D84, D91

[1] corresponding author; Institute for Statistical Studies and Economics of Knowledge, National Research University Higher School of Economics (HSE University), 20 Myasnitskaya St., Moscow, 101000, Russia; and School of Economics and Management, Far Eastern Federal University, 8, Sukhanova St., Vladivostok, 690990, Russia inga.ivanova@hse.ru;
[2] Department of Finance and Risk Management, Warsaw University of Technology, Narbutta 85, 02-524 Warsaw, Poland e-mail: grzegorz.rzadkowski@pw.edu.pl
[3] Amsterdam School of Communication Research (ASCoR), University of Amsterdam PO Box 15793, 1001 NG Amsterdam, The Netherlands; loet@leydesdorff.net. *Loet Leydesdorff contributed to an early version of the paper before his passing away in Mars, 2023*

## I. Introduction

The mechanisms that control the price movements of market assets are of great concern to investors and policymakers. A quick glance at a price chart of a market asset reveals periods of directional price movement and periods of price fluctuations. This chart reflects investors' decisions to buy, sell, or exit the market at specific points in time. Investors, as humans, are guided by logic and psychology in their decisions. The delicate balance between these two factors determines the price of an asset. In short, the price of a market asset is determined by investors' decisions, and decisions are determined by their respective trading preferences. But what drives the preferences? It can be incoming information that is either rationally analyzed or distorted by psychological biases. This analysis provides information with meaning, which is the basis for further action. The question is whether this incoming information is processed in the same way or supplied with different meanings by different groups of investors. This paper attempts to study the mechanisms that drive investor preferences.

The first aspect of the paper concerns the mechanisms of formation of financial time series and the possibility of forecasting future price movements. There are different approaches to studying market dynamics in an attempt to predict future changes. One of the ideas of assessing the evolution of prices of market assets is that past prices can indicate their future values in accordance with observed market trends (e.g. Miner, 2002).

Another approach argues that future price values have no relation to past price values, so that asset prices resemble the movements of molecules (Osborn and Murphy, 1984; Malkiel, 1973). Consequently, market movements cannot be predicted.

The efficient market hypothesis (EMH) (e.g. Black & Scholes, 1973; Gulko, 1997) assume that asset prices reflect all available information, so that future prices cannot be predicted. However, there are also examples of predictable price behavior (e.g. Kendall, 1953), which are considered anomalous under the EMH. Furthermore, machine learning-based studies have been reported to provide high forecasting accuracy for financial time series (e.g. Patel *et al*., 2015; Hsua *et al*., 2016).

Behavioural economics, which focuses on the various psychological factors that influence people's economic decisions, takes an approach that differs from the EMH (e.g., Kahneman and Tversky, 1979; Thaler, 1980, 1985; Banerjee, 1992). It argues that the decisions of economic agents can be largely considered irrational and driven by psychological factors, so that the economic worldview of rational agents can no longer be supported. A number of publications cite studies of the mechanisms that govern investor decisions in financial markets (e.g., Shiller, 1981; Statman, 1995; Olsen, 1998; Barber & Odean, 1999). Because deviations from full rationality are systematic and can be modelled and studied, knowledge of these mechanisms can be used to improve forecasts of future investor behaviour and asset price movements.

For example, in the case of herd behavior, previously formed market trends can be extrapolated to some future period. This indicates that the market is to some extent predictable, and past prices can sometimes be used to predict the direction of price changes (Lo & Mackinlay, 2002). The adaptive market hypothesis (Lo, 2004, 2005) views the EMH and behavioral approaches as opposite sides of the same coin, interpreting the behavioral biases of market participants in an evolutionary aspect as an adaptation to changing conditions of the market environment, so that each group of market participants behaves accordingly in its own way. In its evolutionary aspect, the adaptive market hypothesis is close to the complex systems approach, which views the market as a complex evolving system of interconnected networks of interacting agents (e.g. Krugman, 1995; Arthur, Durlauf, and Lane, 1997; Farmer *et al*., 2012).

While behavioral finance primarily focuses on individual-level biases, social finance, a new emerging paradigm in financial research, focuses on cultural traits including information cues, beliefs, strategies, etc., adopted by larger groups of investors (Hirshleifer, 2015; Akçay and Hirshleifer, 2021).

A common assumption in these approaches is that asset prices reflect investors' reactions to information. But the questions that need to be asked are: 1) how do investors process information, 2) does the same information have the same meaning for different groups of investors, and 3) does the way investors process information matter more than the information itself? The motivation for this paper is to answer these questions.

This paper presents a new approach based on the processing of information by groups of investors that guides their behavior (hereinafter referred to as the information approach). When information is received, it must first be processed, i.e. supplied with meaning. However, information may be processed differently by different groups of investors, which provide different criteria by which information can be given meaning. These criteria can be defined as selection environments in terms of certain coding rules (or sets of communication codes). Coding rules drive latent structures that organize different meanings into structural components (Leydesdorff, 2010). "Meanings emerge from communications and feedback on communications. When choices can influence one another, complex and potentially nonlinear dynamics are generated" (Leydesdorff, 2021, p. 15). Meanings generate expectations about possible states of the system, which are generated relative to future moments. Expectations act as feedback on the current state (i.e., against the arrow of time). In other words, the system simultaneously assumes its past, present, and future states, in accordance with Bachelier's observation that "past, present, and even discounted future events are reflected in the market price" (Bachelier, 1964).

Expectations provide a source of additional options for possible future states of the system that are available but not yet realized. The more options a system has, the more likely it is that the system will deviate from the previous state in the process of autocatalytic self-organization. A measure of additional options is redundancy, which is defined as the addition of information to the maximum information content (Brooks & Wiley, 1986). The concept of redundancy has been applied to innovation research regarding synergies in the Triple Helix (TH) model of university-industry-government relationships (e.g., Etzkowitz & Leydesdorff, 1995, 1998; Leydesdorff, 2003; Park & Leydesdorff, 2010; Leydesdorff & Strand, 2013, etc.). The interaction between differently shaped investors' expectations of different shapes can ultimately generate nonlinear market price dynamics.

The contribution of this paper is twofold: 1) in a narrow sense, it presents a model of market asset price dynamics that can shed light on the mechanisms governing the formation of time series of market asset prices and, in some cases, can predict future price movements; 2) in a broad sense, it is an advance in the theory of meaning that extends this theory to broader practical areas. The first exploratory goal of this paper is to test the applicability of the general concept of information and meaning communication to describe market price dynamics. The second exploratory goal is to provide a quantitative description of market price movements based on the evolutionary dynamics of investor expectations. We use a computationally efficient technique of logistic Continuous Wavelet Transform (CWT) to analyse weekly EUR/USD data for the period 2001.10.07 -2023.08.27. Revealed trend structure coincides with theory predicted patterns. This approach may help practitioners and policymakers to reveal and analyse the latent trend structure in price dynamics and make informed decisions with respect to future asset price movements and market crashes.

## II. Literature Review

The approach that exploits meaning generation in inter-social communications is not well established in the finance literature. However, some basic model characteristics can be found in the literature. Time series financial modelling has been a subject of debate among academics and practitioners. Much of the literature focuses on machine-based time series forecasting. Related model types include traditional machine learning (ML) models (e.g., Bahrammirzaee, 2010;

Mullainathan, S., and Spiess, 2017; etc.) and emerging deep learning (DL) models in the ML field, such as artificial neural networks (ANNs), recurrent neural networks (RNNs), convolutional neural networks (CNNs), and deep multilayer perceptrons (DMLPs) (e.g., Schmidhuber, 2015; Sokolov *et al.*, 2020). The advantage of the analytical approach over the numerical one is that in analytical models it is possible to trace the properties of the time series to the behavior of investors. Frequently observed and difficult to explain facts such as excess volatility (LeRoy & Porter, 1981; Shiller, 1981), volatility clustering (Mandelbrot, 1997), etc. have led to a growing number of works devoted to heterogeneous agent-based models (HAM) (e.g. Hommes, 2006, Chiarella, Dieci, & He, 2009), in which the financial market consists of different groups of agents. The interaction of agents can generate complex market dynamics, including both chaos and stability (e.g. Brock and Hommes, 1997; Bonabeau, 2002). Kaizoji (2004) showed that intermittent chaos in asset price dynamics can be observed in a simple model of financial markets with two groups of agents, which can be explained by the heterogeneity of traders' trading strategies. The behavior of financial markets during crises can be described using rogue waves (e.g., Jenks, 2020). This type of waves is analytically described in a nonlinear option pricing model (Yan, 2010). These solutions can be used to describe possible mechanisms of the rogue wave phenomenon in financial markets. Rogue waves can be found in Korteweg de Vries (KdV) systems if we take into account real non-integrable effects, higher-order nonlinearity, and nonlinear diffusion (Lou & Lin, 2018). Dhesi & Ausloos (2016) observed a kink effect resembling a soliton behavior when studying the behavior of agents reacting to time-dependent logarithm return news in the framework of an irrational fractional Brownian motion model. They also asked the question: what is the differential equation whose solution describes this effect? In some ways, the idea behind this approach is close to that used in opinion dynamics models (e.g., Zha *et al.*, 2020; Granha *et al.*, 2022).

In the absence of structural changes in fundamentals, the period of price explosions has a non-fundamental explanation, i.e., cognitive bias. The inability of investors to properly process available information can cause systematic patterns in price movements, such as seasonality. This inability can be explained by cognitive bias on the part of investors (Hirshleifer, 2015; Fang *et al.*, 2021), who use ease-of-processing heuristics when processing information related to the pricing of market assets. Knowledge of seasonality mechanisms can provide forecasts for the movement of repo rates. Biased beliefs about future price movements are an important driver of market prices. Changes in investors' expectations about future stock market returns can explain the facts about stock market price movements (De la O and Myers, 2021). Jin and Sui (2022) showed that investors' beliefs about future market returns can generate excess volatility in stock market returns. These beliefs are largely dependent on recent past returns and can provide significant predictability of returns. Group behavior of investors in interpreting external information also affects stock price dynamics. Massa, O'Donovan, and Zhang (2021) showed that strategic risk reallocation by business groups in response to news that can be interpreted as "bad" information is an additional determinant of stock returns. Duffy, Rabanal, and Rud (2021) studied how exchange-traded funds (ETFs) affect asset pricing and turnover in a laboratory asset market consisting of three subgroups of traders, where each subgroup follows its own preferred trading strategy. The three-component market structure is also described in the SIR model, which is used to explain investor idea diffusion and bubbles (Shiller, 2019). The SIR model, in turn, can be linked to the TH model and the corresponding nonlinear dynamics described by the KdV equation can be observed (Ivanova, 2022). Noussair and Popesku (2021) provide evidence that co-movement of market assets in the absence of a common shock to underlying fundamentals can be explained by behavioral factors caused by asymmetric information between informed and uninformed traders. Co-movement can occur between two risky assets even if their fundamentals are uncorrelated, and after a dividend shock, the shocked asset exhibits autocorrelation in returns. A new paradigm for understanding financial markets is social finance (Akçay and Hirshleifer, 2021). Social finance argues that the behavior of financial market participants depends on accepted cultural traits that determine the patterns of market operation. These traits are shared by larger groups of investors and provide informational transmission distortions between investor groups. Investors' financial traits are unstable and can change over time.

### III. Theory

The theory of meaning generation in inter-societal communication (e.g., Leydesdorff, 2021) is based on an information-theoretic approach to measuring redundancy as an indicator of synergy in the Triple Helix model of innovation. The market can be viewed as an autocatalytic system whose complex dynamics determine the movement of market prices. The market consists of different groups of investors - pension funds, banks, hedge funds, public corporations, individuals, etc. - that behave according to their investment preferences.

Investors, with respect to their sentiments, can be divided into three large groups (agents), which include investors who expect prices to rise, fall, and the group of indecisive investors, waiting for more favorable conditions to enter the market. Agents make their decisions based on the information they receive. This information must first be provided with meaning (as a "signal") or discarded (as noise). Each group uses different criteria to filter the information according to its behavioral biases regarding its market positions. That is, the same information is viewed from different perspectives (or positions) and can be provided with different meanings. The mechanisms of information and meaning processing are different. While information can be transmitted through a network of relations, meanings are provided from different positions (Burt, 1982). Meaning cannot be communicated, but can only shared when positions overlap. Meaning processing can increase or decrease the number of options, which can be measured as redundancy (Leydesdorff & Ivanova, 2014). The redundancy calculus is complementary to the calculus of information. Shannon (1948) defined information as a probabilistic entropy: $H = -\sum_i p_i \log p_i$, which is always positive and adds to the uncertainty (Krippendorff, 2009). Two overlapping distributions with information content $H_1$ and $H_2$ can be considered (Figure 1).

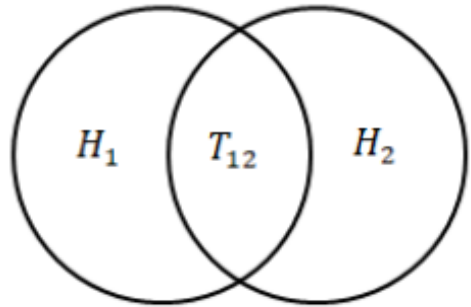

**Figure 1**: Set-theoretical representation of two overlapping distributions with informational contents $H_1$ and $H_2$

Total distribution is the sum of two distributions minus overlapping area, since it is counted twice:

$$H_{12} = H_1 + H_2 - T_{12} \qquad (1)$$

Overlapping area relates to mutual or configurational (McGill, 1954) information ($T_{12}$). The formula (1) can be written as:

$$T_{12} = H_1 + H_2 - H_{12} \qquad (2)$$

In case of three overlapping distributions (Figure 2)

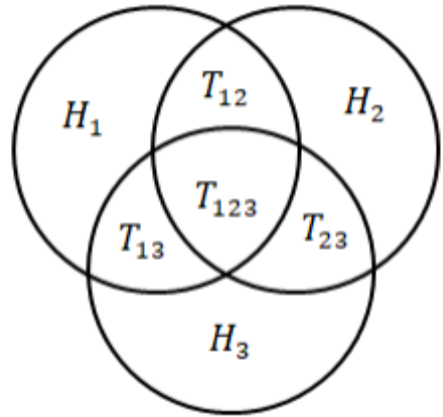

**Figure 2:** Set-theoretical representation of three overlapping distributions with informational contents: $H_1$, $H_2$ and $H_3$

Configurational information (e.g., Abramson, 1963) is:

$$T_{123} = H_1 + H_2 + H_3 - H_{12} - H_{13} - H_{23} + H_{123} \qquad (3)$$

However, $T_{123}$ is no longer Shannon-type information, since it can be negative. The sign changes with each new distribution added (e.g., Krippendorff, 2009). Technically, the sign change problem can be solved by introducing "positive overlap" (Leydesdorff and Ivanova, 2014). This time, no allowance is made for the overlap, which is counted twice and therefore redundant, but other mechanisms are assumed by which the two distributions influence each other. These mechanisms are different from the relational exchange of information and lead to an increase in redundancy, so that the overlapping region is added rather than subtracted. In formula format, this can be written as:

$$H_{12} = H_1 + H_2 + R_{12} \qquad (4)$$

It follows that $R_{12}$ is negative ($R_{12} = -T_{12}$)) and therefore is redundancy - uncertainty reduction (similar to: $R_{123} = T_{123}$). That is, when measuring configurational information in three (or more) dimensions, it is not Shannon-type information that is measured, but mutual redundancy. Since this measure yields a negative amount of information, it can be considered an indicator of synergy between the three sources of variance (Leydesdorff, 2008).

The three investor groups communicate with each other and form a relational network. But there is another mechanism on top of the structural network that governs the evolution of the system. The communicated information is processed differently by each investor group according to different sets of coding rules (communication codes), so that the groups are (positionally) differentiated with respect to their positions toward processing the information. The same information may be provided with different meanings and, depending on investor sentiment, may be taken as a buy or sell signal by different agents. The sets of communication codes are latent but may be partially correlated by forming a correlation network on top of the relational one. Meaning generation is provided from the perspective of hindsight, i.e. what information can mean for future events. Structural differences between coding and decoding algorithms provide a source of additional options in reflexive and anticipatory communication, i.e. meaning-generating structures act as selection environments (Leydesdorff, 2021). These additional options are a source of variation.

Changes occur when a market eventually changes from its previous state. When there are two selection environments, they can shape each other in coevolution and stabilize on a historical trajectory. In the case of three (or more) selection environments, every third environment can disturb the interaction between the other two, so that the three bi-lateral trajectories can shape a three-lateral regime. This mechanism is known as "triadic closure" and governs the evolutionary dynamics of the system (Granovetter, 1973; Bianconi *et al*., 2014). Simmel (1902) pointed out the qualitative difference between dyads and triads. Triads can be transitive or cyclical (Batagelj

*et al.*, 2014). Two cycles can emerge – positive (autocatalytic) and negative (stabilizing) (Figure 3). The autocatalytic cycle enhances the change in the previous state of the system (the system self-organizes), while the stabilizing cycle keeps the system from transforming.

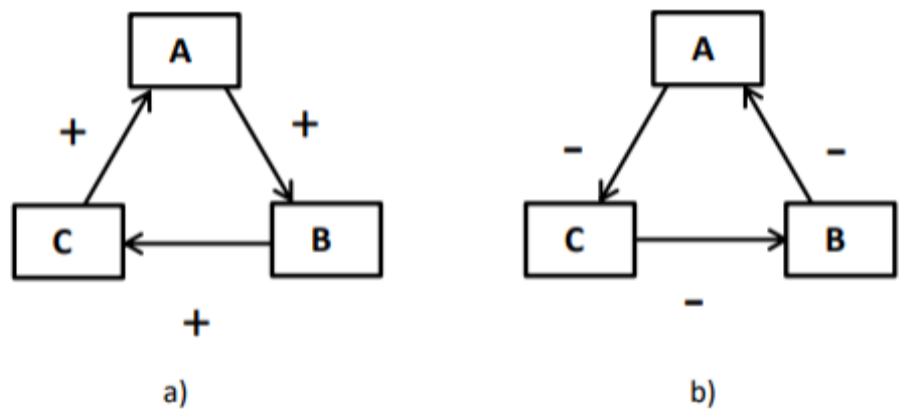

**Figure 3**: Schematic of three-component positive a) and negative b) cycles (Adapted from Ulanovitz, 2009)

The dynamics of information and meaning can be assessed empirically using the sign of mutual information (*R*) as an indicator. The balance between historical stabilization and evolutionary self-organization can be described by the formula (Ivanova and Leydesdorff, 2014a, 2014b)

$$R \sim P^2 - Q^2 \tag{5}$$

The first term in Eq. 5 refers to the positive entropy generated as a result of historical stabilization, while the second term corresponds to the negative entropy generated at the regime level as a result of self-organization processes. Historical stabilization refers to historically realized options that are generated through the recursive regime, while self-organization relies on new, not yet realized options generated through the incursive regime. The trade-off between historical realization and self-organization leads to redundancy cyclical evolution. Dubois (2019) shows that for temporary cyclical systems, probabilities $p_i$ can oscillate around their mean values $p_{i0}$ in a harmonic or non-harmonic mode.

For non-harmonic oscillations, we obtain (see Appendix A):

$$\frac{1}{k}\frac{d^2 p_i}{dt^2} = -(p_i - p_{i0}) + \alpha(p_i - p_{i0})^2 + C_i \tag{6}$$

The probability density function $P$ satisfies the non-linear evolutionary equation (see Appendix B for the derivation):

$$P_T + 6PP_X + P_{XXX} + C_1 = 0 \tag{7}$$

which is the generalization of the well-known Korteweg-de Vries equation:

$$U_T + UU_X + U_{XXX} = 0 \tag{8}$$

Eq. 7 admits soliton solutions. A single soliton solution has the form:

$$P(X,T) = 2\left(\frac{\kappa}{2}\right)^2 ch^{-2}\left[\frac{\kappa}{2}\left(X - 4\left(\frac{\kappa}{2}\right)^2 T + \frac{C_1}{2}T^2\right)\right] - C_1 T \tag{9}$$

Solution of equation 8 can be written in a more general form (here we set $= \frac{\kappa}{2}$):

$$P(X,T) = n(n+1)\rho^2 ch^{-2}\left[\rho\left(X - 4\rho^2 T + \frac{C_1}{2}T^2\right)\right] - C_1 T \tag{10}$$

The pulse in the form of Eq. 8 eventually evolves into a sequence of n single waves with amplitudes: $2\kappa^2, 8\kappa^2, 18\kappa^2 \ldots 2n^2\kappa^2$ and the corresponding velocities $4\kappa^2, 16\kappa^2, 32\kappa^2, \ldots 4n^2\kappa^2$ (Miura, 1976).

There is also a direct method for finding soliton solutions to Eq. 8 that involve multiple solitons with arbitrary amplitudes, so that the *N*-soliton solution takes the form:

$$P = 2\frac{d^2}{dX^2}logF_N \tag{11}$$

where:

$$F_N = \sum_{\mu=0,1} exp\left(\sum_{i=1}^{N} \mu_i \eta_i + \sum_{1 \le i < j}^{N} \mu_i \mu_j A_{ij}\right) \tag{12}$$

Here $\eta_i = k_i X - k_i^3 T$; $A_{ij}$ are the phase shifts of the solitons: $e^{A_{ij}} = \left(\frac{k_i - k_j}{k_i + k_j}\right)^2$ (Ablowitz and Segur, 1981)[4]. It follows from Eq. 10 that the corresponding *N*-soliton solution for Eq. 5 is:

$$\Phi_N = exp\left[-\frac{C}{2}tx^2 + Ax + B\right] \cdot \sum_{\mu=0,1} exp\left(\sum_{i=1}^{N} \mu_i \eta_i + \sum_{1 \le i < j}^{N} \mu_i \mu_j A_{ij}\right) \tag{13}$$

The additional term on the right-hand side of Eq. 5 adjusts the soliton amplitude over time. In the case of a sequence of solitons, there is a relationship between the soliton amplitudes and the time intervals: $\frac{A_i - A_j}{T_i - T_j} = const$. Under the weakly nonlinear assumption, the equation for the redundancy *R* is similar to Equation 7 (Appendix B):

$$4R_T - 2RR_X + R_{XXX} + C_1 = 0 \tag{14}$$

There are also periodic solutions to Eq. 6 (Appendix C, see Lax, 1974).

---

[4] The sum over $\mu = 0,1$ refers to each of $\mu_i$. E.g. performing the calculation for N=3 yields $F_3 = 1 + e^{\eta_1} + e^{\eta_2} + e^{\eta_3} + e^{\eta_1 + \eta_2 + A_{12}} + e^{\eta_1 + \eta_3 + A_{13}} + e^{\eta_2 + \eta_3 + A_{23}} + e^{\eta_1 + \eta_2 + \eta_3 + A_{12} + A_{13} + A_{23}}$

The transmitted information is processed using communication codes, and expectations about the future time are generated at the system level. These expectations can be thought of as a density of redundancy representing unrealized but possible options distributed over a time interval. Here, expectations are analytical events (options), and actions are historical events,[5] that can be observed over time as a response to expectations (as if expectations move against the arrow of time and turn into actions). In other words, there is a dynamics of action in historical events below and a dynamics of expectation above that acts reflexively. Expectations are eventually transformed into actions and represent new of the system[6].

Initial expectations arise as a set of market beliefs. These expectations are projected into the future and further stratified according to the nonlinear dynamics of information processing by investors. Finally, expectations are realized and transformed into observable changes in the prices of market assets, forming certain wave patterns along the t-axes. The described mechanism is expected to operate across all price and time scales, creating a self-similar fractal structure.

## IV. Data

The paper relies on publicly available data sets. The data comprise EUR/USD weekly data for the period 07/10/2001 – 27/08/2023. The choice of the EUR/USD pair is due to the fact that EUR, as a currency distributed over many countries, can be expected to demonstrate more stable dynamics with fewer random fluctuations than single country currency.

---

[5] Shannon (1948) defined the proportion of non-realized but possible options as redundancy, and the proportion of realized options as the relative uncertainty or information.

[6] According to the second law of thermodynamics a system's entropy increase with time. For isolated systems it can reach thermodynamic equilibrium

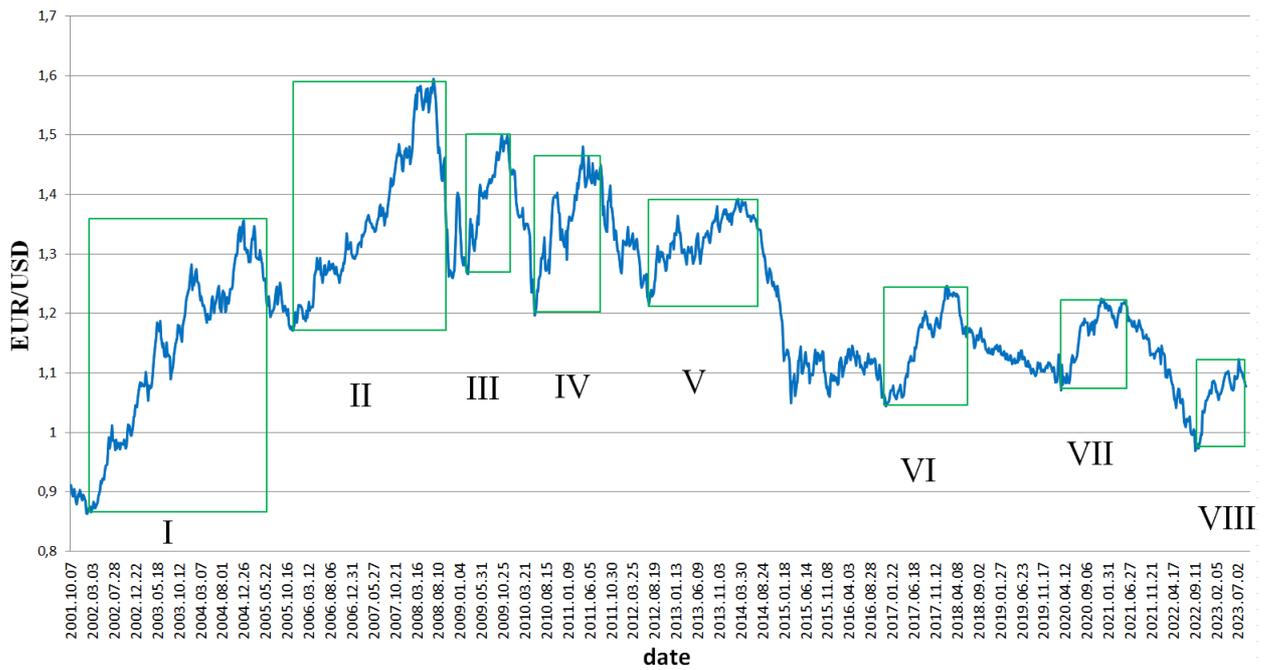

**Figure 4:** EUR/USD weekly trend

We have numbered the weeks consecutively from 1 (07/10/2001) to 1143 (27/08/2023). The periods of growth of the EUR/USD index I-VIII occurred in the following weeks, Table 1

| Period | Start of period (week) | End of period (week) | Length of period (weeks) |
|---|---|---|---|
| I | 1 (07/10/2001) | 196 (03/07/2005) | 196 |
| II | 216 (20/11/2005) | 361 (31/08/2008) | 146 |
| III | 387 (01/03/2009) | 428 (13/12/2009) | 42 |
| IV | 452 (10/05/2010) | 518 (04/09/2011) | 67 |
| V | 564 (22/07/2012) | 667 (13/07/2014) | 104 |
| VI | 793 (11/12/2016) | 865 (29/04/2018) | 73 |
| VII | 971 (10/05/2020) | 1029 (20/06/2021) | 59 |
| VIII | 1095 (25/09/2022) | 1140 (06/08/2023) | 46 |

**Table 1**. Periods of growth of the index EUR/USD

## V.     Method

Under the assumption that the evolution of redundancy is described by Eq. (14) it is natural to look for an expansion of the longitudinal currency time series in the form of a sequence of solitary waves. For the purpose we develop a method for time series decomposition based on wavelet analysis. Wavelet analysis is widely used in economics and finance to study co-movement among foreign exchange markets, extract the main signals, locate the discontinuities in the data and forecast exchange rates (e.g. Yang *et al*., 2016; Nguyen & He, 2015; Jin, J. and Kim, J. 2015). The advantage of using wavelet analysis in comparison with e.g. Fourier transform is that it doesn't require strong assumptions about the data generating processes and allows decompose highly volatile and complex financial data in a sum of just few waves.

In general, there is no preference regarding the choice of wavelet basis, and the results obtained, although they reflect the dynamics of the financial system in a simplified form, can hardly be conceptualized within a theoretical framework. A distinctive feature of our approach is that instead of directly analyzing the given time series using the wavelet transform, we: 1) generate an aggregate time series from a given time series; 2) decompose aggregate time series into logistic components using continuous wavelet transform (CWT) applied to the second differences of the aggregate time series[7] and determine the parameters of logistic waves from CWT scalogram[8]; 3) determine the parameters of the derivatives of logistic waves, with the help of which the given time series are analytically approximated. Thus, the time series are approximated by the sum of derivatives of logistic waves, which have the form of soliton solutions of the KdV equation. Obviously, instead of calculating second differences of the aggregated time series, one can apply CWT to the first differences of the given time series. The resulting scalogram will not change.

---

[7] This decomposition can be treated as a complement to the paper by Meyer *et al*., (1999)
[8] A scalogram is a three-dimensional graph where the Z axis is indicated by a colour that represents CWT values, and the X and Y axes respectively provide time and scale information. The possibility of retrieving the parameters of original function from the scalogram is confirmed by Mallat and Hwang (1992).

Next we will briefly present the basic information from the theory of wavelet transforms.

A (mother) wavelet (Daubechies, 1992, p.24) is an integrable function $\psi \in L^1(\mathbb{R})$, which satisfies the following admissibility condition:

$$C_\psi = 2\pi \int_{-\infty}^{\infty} |\xi|^{-1} |\hat{\psi}(\xi)|^2 d\xi < \infty \tag{15}$$

where $\hat{\psi}(\xi)$ is the Fourier transform of $\psi$

$$\hat{\psi}(\xi) = \frac{1}{\sqrt{2\pi}} \int_{-\infty}^{\infty} \psi(x) e^{-i\xi x} dx$$

Moreover, we assume that $\psi$ is square integrable, with the norm:

$$\|\psi\| = \|\psi\|_{L^2} = \left( \int_{-\infty}^{\infty} |\psi(x)|^2 dx \right)^{1/2}$$

Dilating and translating the mother wavelet one obtains a family of wavelets:

$$\psi^{a,b}(x) = \frac{1}{\sqrt{a}} \psi\left( \frac{x-b}{a} \right)$$

where $a, b \in \mathbb{R}$, $a > 0$ and $\|\psi^{a,b}\| = \|\psi\|$. Usually, $\|\psi\| = 1$. The value of the Continuous Wavelet Transform (CWT) at a point $(a, b)$ of a function $f \in L^2(\mathbb{R})$ is the inner product of the function $f$ and the wavelet $\psi^{a,b}$:

$$(T^{wav} f)(a, b) = \langle f, \psi^{a,b} \rangle = \int_{-\infty}^{\infty} f(x) \psi^{a,b}(x) dx \tag{16}$$

where $a, b$ are parameters of wavelet family.

Here we use normalized second order logistic wavelets. Logistic mother wavelet of order $n$ in an unnormalized form is defined as the $n$th derivative $x^{(n)}(t)$ of the basic logistic function $x(t) = \frac{1}{1+e^{-t}}$ (Rzadkowski and Figlia, 2021). However, in order to be able to compare different wavelets, it is convenient to deal with a normalized family of wavelets.

One can show that (Appendix D):

$$\int_{-\infty}^{\infty}\left(x^{(n)}(t)\right)^2 dt = (-1)^{n-1} B_{2n} = |B_{2n}| \tag{17}$$

where $B_{2n}$ is the $(2n)$th Bernoulli number. So the logistic mother wavelet $\psi_n(t)$ of order $n$ ($n = 2, 3, ...$) defined as:

$$\psi_n(t) = \frac{(-1)^n}{\sqrt{|B_{2n}|}} x^{(n)}(t) \tag{18}$$

is now normalized to unity $\|\psi_n\| = \|\psi_n\|_{L^2} = 1$.

Since $B_4 = -1/30$, then from (18) we see that, in particular, the normalized second order mother logistic wavelet has the form:

$$\psi_2(t) = \sqrt{30}\, x''(t) = \frac{\sqrt{30}(e^{-2t} - e^{-t})}{(1+e^{-t})^3} \tag{19}$$

and $\psi_2^{a,b}(t) = \frac{1}{\sqrt{a}} \psi_2\left(\frac{t-b}{a}\right)$.

Suppose we want to estimate unknown parameters $x_{sat}$-saturation level, $b$ - inflection point and $a$-slope coefficient of a logistic function $\frac{x_{sat}}{1+\exp\left(-\frac{t-b}{a}\right)}$. More generally let us consider a combination of the logistic function and a linear function with real constants $c$ and $d$:

$$f(t) = c + dt + \frac{x_{sat}}{1 + \exp\left(-\frac{t-b}{a}\right)}$$

In Appendix E we show that the logistic CWT applied to the second derivative $f''(t)$ of this function

$$(T^{wav} f'')(\alpha, \beta) = \langle f'', \psi_2^{\alpha,\beta} \rangle = \int_{-\infty}^{\infty} f''(t)\, \psi_2^{\alpha,\beta}(t) dt$$

takes maximum (for $x_{sat} > 0$) or minimum (for $x_{sat} < 0$) value when $\alpha = a$ and $\beta = b$. Therefore, knowing the maximum (or minimum) value of the CWT we can read the values of parameters *a* and *b* of the function $f(t)$ directly from the scalogram and we can calculate the saturation level of it (Eq. E1). For increasing logistic function ($x_{sat} > 0$), we get

$$x_{sat} = \sqrt{30}\, a^{3/2} \max(T^{wav} f'') \qquad (20)$$

and for decreasing logistic function ($x_{sat} < 0$)

$$x_{sat} = \sqrt{30}\, a^{3/2} \min(T^{wav} f'') \qquad (21)$$

### VI. Algorithmic formalization

For a given time series $(y_n)$, $n = 0,1,2,...,N + 1$ we define the central second differences

$$\Delta^2 y_n = (y_{n+1} - 2y_n + y_{n-1}), \qquad n = 1,2,3,...,N$$

In supposition that the time series $(y_n)$ follows the logistic trend: $y_n = y(n) = \frac{y_{sat}}{1+\exp\left(-\frac{n-b}{a}\right)}$ we apply Matlab's CWT to $\Delta^2 y_n$. Then for a specific range of parameters $\alpha, \beta$ the Matlab's command cwt returns the value of the following sum (Index):

$$\text{Index} := \sum_{n=1}^{N} \Delta^2 y_n \psi_2^{\alpha,\beta}(n) \approx \int_{-\infty}^{\infty} y''(t)\, \psi_2^{\alpha,\beta}(t)\, dt$$

In the above equation we use the well-known fact in numerical analysis, that the second difference $\Delta^2 y_n$ can be used to approximate the value of the second derivative $y''(n)$. A more detailed explanation of this fact for logistic wavelets is analogous to what was done in Rzadkowski (2024, Remark 4.1, p.11) in the case of the Gompertz wavelets.

Reading from the CWT scalogram, the values of parameters $\alpha, \beta$, for which the Index takes the maximum or minimum value, estimating the values of parameters $\alpha \approx a, \beta \approx b$ of the initial logistic wave $y(t)$, and then estimating its saturation level by using (20) and (21):

$$y_{sat} \approx \sqrt{30}\, \alpha^{3/2} \max(\text{Index}) \quad \text{or} \quad y_{sat} \approx \sqrt{30}\, \alpha^{3/2} \min(\text{Index}) \tag{22}$$

we can model the time series $(y_n)$ by using an adequate logistic function.

Similarly we can do in the case of a multilogistic function (plus a linear trend)

$$y(t) = ct + d + \sum_{i=1}^{k} \frac{y_{i,sat}}{1+\exp\left(-\frac{t-b_i}{a_i}\right)} \tag{23}$$

$i = 1,2,...,k$, where $k$ is the number of logistic waves. If there are several overlapping logistic waves, occuring in the same time period, then the higher intensity waves (with a higher Index) may cause lower-intensity waves to be invisible on the CWT scalogram. Therefore, in order to find such waves, we can remove the first wave with the highest intensity by subtracting it from the time series $(y_n)$:

$$y_n^{(1)} = y_n - \frac{y_{1,sat}}{1 + \exp\left(-\frac{t-b_1}{a_1}\right)}$$

Then, for the time series $\left(y_n^{(1)}\right)$, we calculate its central second differences and for the latter we perform the CWT analysis again. The above process may be repeated several times if necessary. Usually, in order to more precisely estimate the values of some parameters of the multilogistic function, we can use some optimization methods.

### VII.   Results

The model suggests that when nonlinear effects dominate, asset prices evolve in trends that can be described by the nonlinear evolutionary equation (7). To answer the question of whether the concept of information-semantic communication can be applied to the description of market price dynamics, one can try to find patterns that can be described by the model. We compare the model results with empirically observed weekly EUR/USD data. The same patterns can be found in different markets and (due to the fractal nature of stock and financial markets) in different time frames.

We performed the calculation separately for eight ascending trends labeled in Fig.4 by roman numbers I - VIII. First we calculate the CWT for large values of parameter *a* and find the largest waves. Then we find the parameters of the medium and small waves (Fig. 5). If needed, we also use some optimization methods (local minimization of the RMSE error). Numbers indicate corresponding trends in Fig.5.

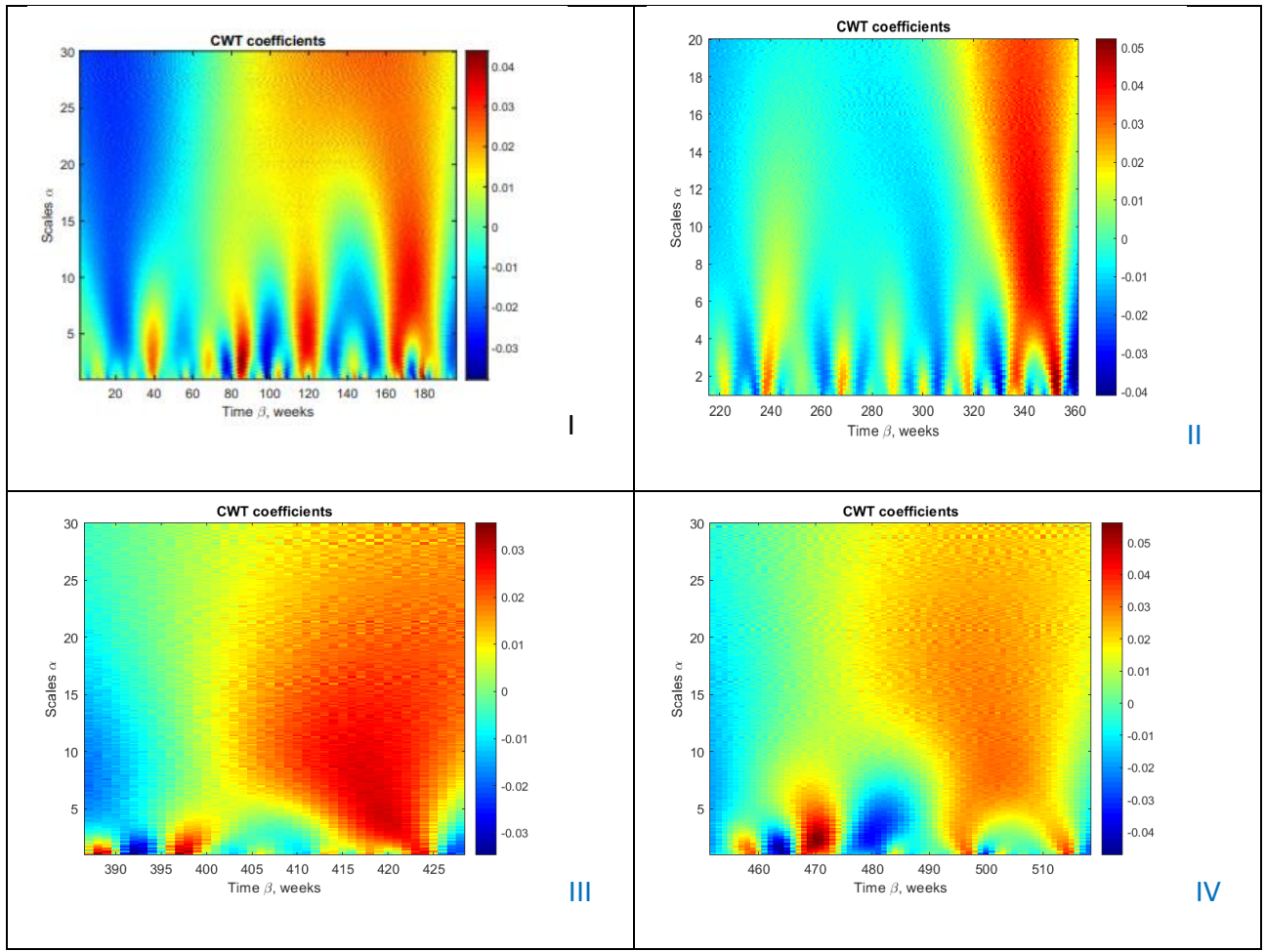

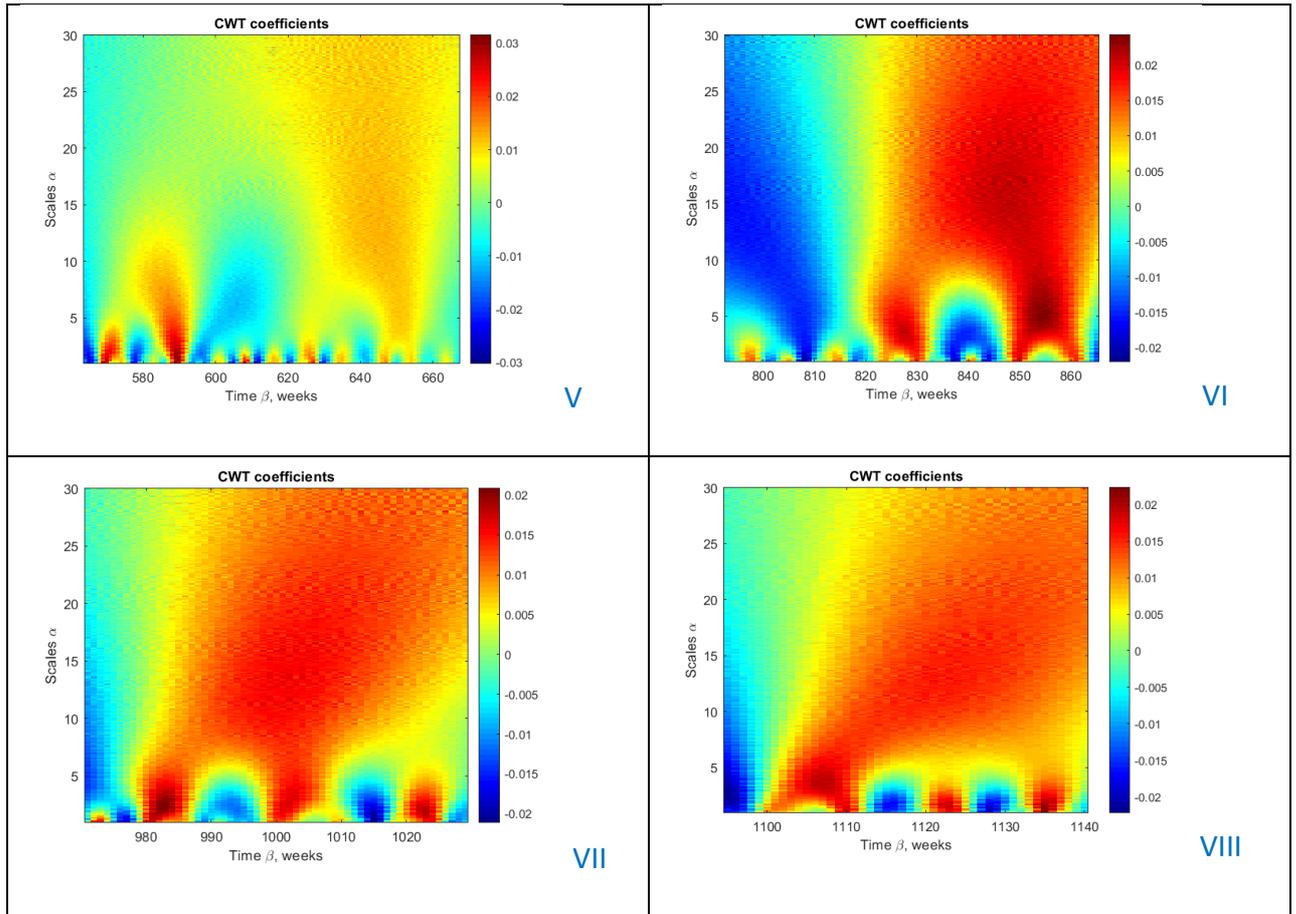

**Figure 5:** Scalograms for periods I - VIII

In this way, we find the parameters of logistic waves approximating our data, Table 2.

**Table 2:** Waves in periods I - VIII

| $i$ | $a_i$ | $b_i$ | $y_{i,sat}$ | ratio |
|---|---|---|---|---|
| 1 | 15.43 | 20.5 | -9.39 | -0.00742 |
| 2 | 3.64 | 39 | 0.96 | 0.00169 |
| 3 | 2.88 | 68 | 0.68 | 0.000868 |
| 4 | 3.43 | 86 | 1.6 | 0.00136 |
| 5 | 6.61 | 119 | 3.65 | 0.00116 |
| 6 | 6.75 | 132 | -1.41 | -0.000396 |
| 7 | 2.66 | 154 | -0.72 | -0.000439 |
| 8 | 26.82 | 173 | 32.72 | 0.00176 |
| 9 | 3.14 | 194 | -0.95 | -0.00039 |

I

| $i$ | $a_i$ | $b_i$ | $y_{i,sat}$ | ratio |
|---|---|---|---|---|
| 1 | 12.83 | 219 | -5.87 | -0.0191 |
| 2 | 2.16 | 233 | -0.253 | -0.00146 |
| 3 | 2.43 | 239 | 0.347 | 0.001373 |
| 4 | 2.38 | 261 | -0.322 | -0.0007 |
| 5 | 0.828 | 268 | 0.0648 | 0.000356 |
| 6 | 1.83 | 288 | 0.197 | 0.000359 |
| 7 | 2.36 | 318 | 0.359 | 0.000362 |
| 8 | 4.09 | 329 | -0.721 | -0.00038 |
| 9 | 1.46 | 337 | 0.256 | 0.000354 |
| 10 | 15.06 | 344 | 15.33 | 0.00194 |
| 11 | 1.52 | 353 | 0.303 | 0.000356 |
| 12 | 1.4 | 360 | -0.28 | -0.00034 |

II

| $i$ | $a_i$ | $b_i$ | $y_{i,sat}$ | ratio |
|---|---|---|---|---|
| 1 | 6 | 387 | -1.8 | -0.025 |
| 2 | 1 | 390 | 0.16 | 0.00667 |
| 3 | 1.3 | 394 | -0.3 | -0.00577 |
| 4 | 2.7 | 398 | 0.7 | 0.00463 |
| 5 | 1 | 408 | 0.1 | 0.00104 |
| 6 | 1.2 | 414 | 0.1 | 0.000694 |
| 7 | 1.1 | 419 | 0.11 | 0.000714 |
| 8 | 4.6 | 420 | 1.57 | 0.00237 |
| 9 | 1.4 | 425 | 0.3 | 0.00131 |

III

| $i$ | $a_i$ | $b_i$ | $y_{i,sat}$ | ratio |
|---|---|---|---|---|
| 1 | 1.8 | 459 | 0.58 | 0.00895 |
| 2 | 3.06 | 463 | -1.32 | -0.00829 |
| 3 | 5.99 | 470 | 3.83 | 0.00799 |
| 4 | 4.47 | 481 | -2.18 | -0.00393 |
| 5 | 1.13 | 496 | 0.079 | 0.000380 |
| 6 | 20.56 | 500 | 31.08 | 0.00756 |
| 7 | 1.89 | 514 | 0.32 | 0.000661 |

IV

| $i$ | $a_i$ | $b_i$ | $y_{i,sat}$ | ratio |
|---|---|---|---|---|
| 1 | 2.78 | 566 | -0.95 | -0.0285 |
| 2 | 3 | 572 | 0.35 | 0.00324 |
| 3 | 2 | 579 | -0.3 | -0.00234 |
| 4 | 2.1 | 590 | 0.4 | 0.00176 |
| 5 | 1.1 | 599 | -0.14 | -0.000884 |
| 6 | 1.3 | 601 | 0.07 | 0.000354 |
| 7 | 1.2 | 606 | -0.15 | -0.000727 |
| 8 | 1.1 | 610 | 0.13 | 0.000629 |
| 9 | 1 | 613 | -0.16 | -0.0008 |
| 10 | 1.1 | 619 | 0.06 | 0.000244 |
| 11 | 1.2 | 622 | -0.09 | -0.000318 |
| 12 | 1.3 | 628 | 0.17 | 0.000503 |
| 13 | 1 | 631 | -0.13 | -0.000478 |
| 14 | 1.7 | 636 | 0.09 | 0.000181 |
| 15 | 1.3 | 642 | -0.11 | -0.000267 |
| 16 | 1.1 | 649 | 0.08 | 0.000211 |
| 17 | 12.9 | 649 | 3.88 | 0.000874 |
| 18 | 1 | 651 | -0.07 | -0.000199 |
| 19 | 1.1 | 654 | 0.07 | 0.000175 |

V

| $i$ | $a_i$ | $b_i$ | $y_{i,sat}$ | ratio |
|---|---|---|---|---|
| 1 | 12 | 794 | -4 | -0.0417 |
| 2 | 1.4 | 799 | 0.17 | 0.00434 |
| 3 | 1.1 | 802 | -0.05 | -0.00114 |
| 4 | 1 | 806 | 0.06 | 0.00107 |
| 5 | 1.1 | 809 | -0.1 | -0.00134 |
| 6 | 1 | 816 | 0.05 | 0.000521 |
| 7 | 1.1 | 819 | -0.06 | -0.000505 |
| 8 | 2.8 | 828 | 0.35 | 0.000868 |
| 9 | 1.4 | 839 | -0.22 | -0.000836 |
| 10 | 1 | 842 | 0.04 | 0.0002 |
| 11 | 1 | 844.5 | -0.14 | -0.000667 |
| 12 | 15 | 849 | 6 | 0.00175 |
| 13 | 2 | 852.5 | 0.24 | 0.000496 |
| 14 | 1.7 | 861 | 0.24 | 0.000512 |

VI

| $i$ | $a_i$ | $b_i$ | $y_{i,sat}$ | ratio |
|---|---|---|---|---|
| 1 | 4 | 971 | -0.7 | -0.0146 |
| 2 | 1.1 | 974 | 0.09 | 0.00341 |
| 3 | 1.2 | 977 | -0.07 | -0.00162 |
| 4 | 2.5 | 985 | 0.3 | 0.00176 |
| 5 | 1 | 989 | -0.06 | -0.000714 |
| 6 | 1.6 | 995 | -0.15 | -0.000868 |
| 7 | 9 | 1005 | 3 | 0.00225 |
| 8 | 1.5 | 1017 | -0.15 | -0.00051 |
| 9 | 2.7 | 1024 | 0.58 | 0.000959 |

VII

| $i$ | $a_i$ | $b_i$ | $y_{i,sat}$ | ratio |
|---|---|---|---|---|
| 1 | 3 | 1097 | -0.7 | -0.0194 |
| 2 | 1.07 | 1102 | 0.07 | 0.00204 |
| 3 | 3.79 | 1111 | 0.58 | 0.00225 |
| 4 | 2 | 1116 | -0.2 | -0.00114 |
| 5 | 1.9 | 1125 | 0.33 | 0.0014 |
| 6 | 1.9 | 1128 | -0.2 | -0.000774 |
| 7 | 17 | 1133 | 5.04 | 0.0019 |
| 8 | 2.1 | 1136 | 0.3 | 0.00085 |

VIII

To approximate the Total data, we use the multilogistic function

$$y(t) = c + dt + \sum_{i=1}^{k} \frac{y_{i,sat}}{1+\exp\left(-\frac{t-b_i}{a_i}\right)}$$

and to approximate the differential data, we use the derivative of this function

$$y'(t) = d + \sum_{i=1}^{k} \frac{y_{i,sat}\exp\left(-\frac{t-b_i}{a_i}\right)}{a_i\left(1+\exp\left(-\frac{t-b_i}{a_i}\right)\right)^2}$$

The differential data and the approximating function $y'(t)$ are shown in Fig. 6.

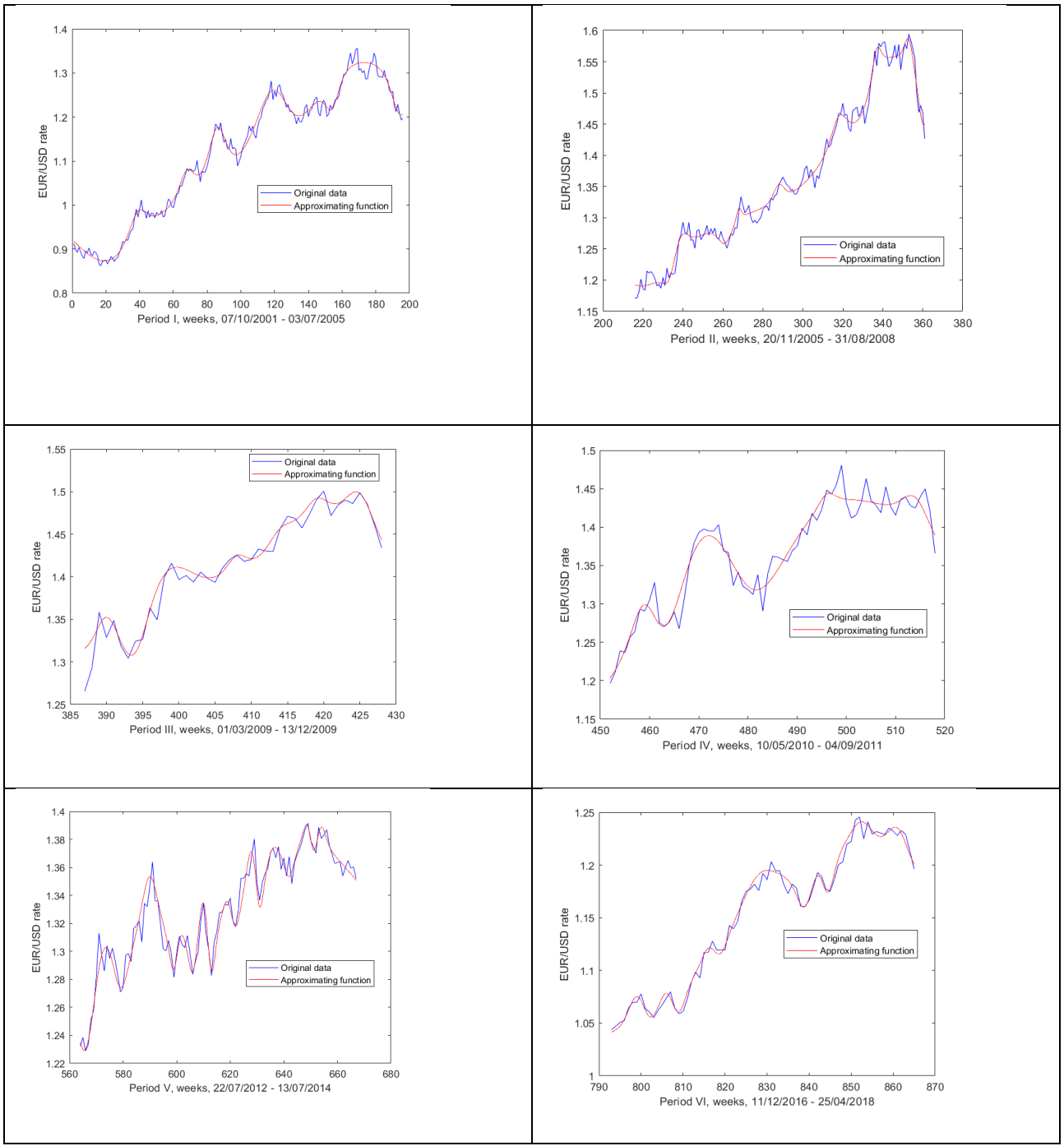

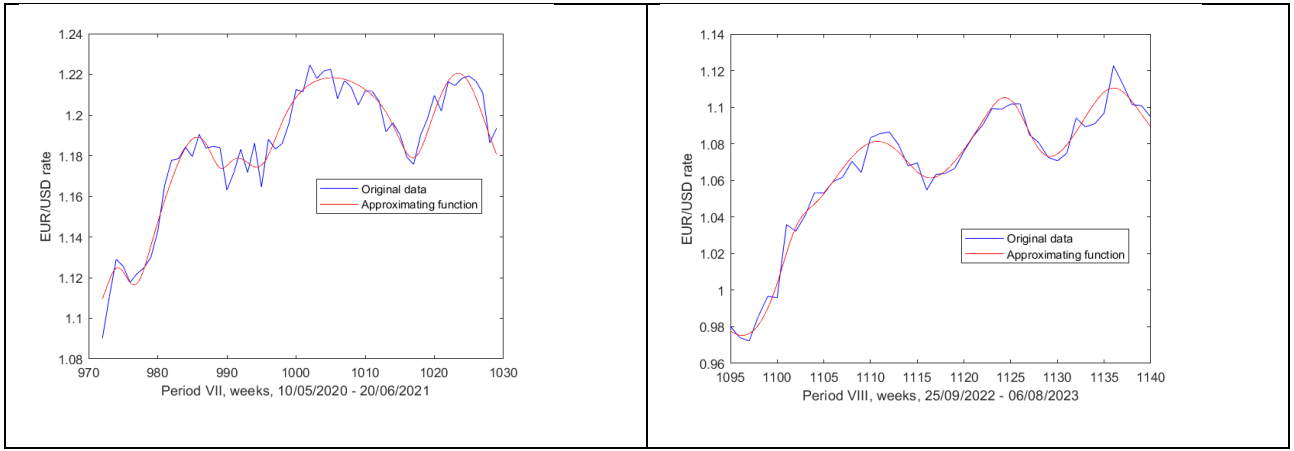

**Figure 6:** EUR/USD rate and approximating function in periods I - VIII

Table 3 gives the values of the coefficient of determination $R^2$ for each period I – VIII.

| Period | $R^2$ coefficient | Period | $R^2$ coefficient |
|---|---|---|---|
| I | 0.991761 | V | 0.956426 |
| II | 0.988207 | VI | 0.992061 |
| III | 0.957631 | VII | 0.955061 |
| IV | 0.943458 | VIII | 0.976349 |

**Table 3**. $R^2$ coefficient for periods I - VIII

Wavelet decomposition of the above periods is shown in Fig. 7

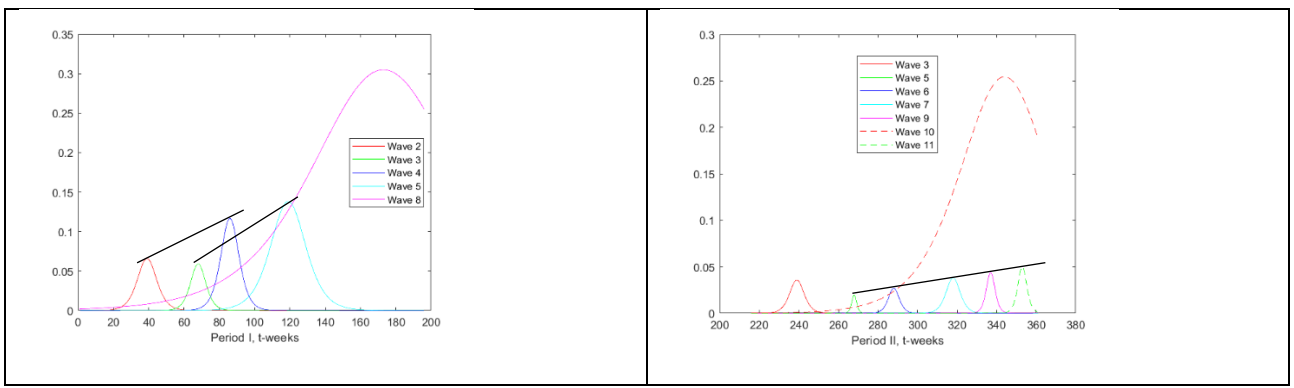

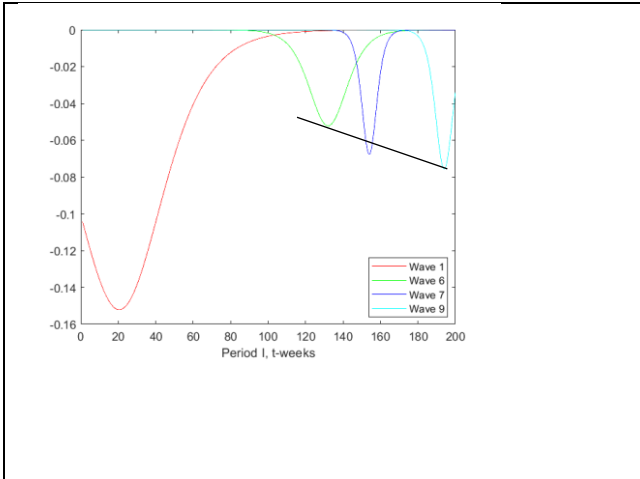
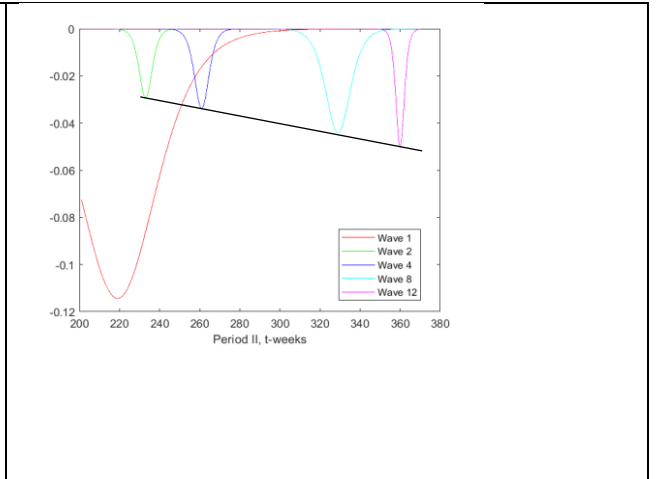
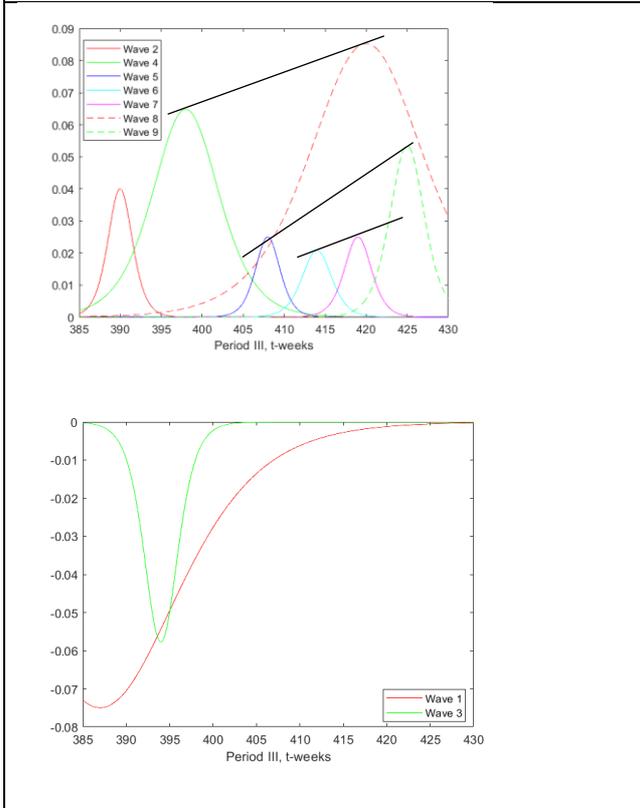
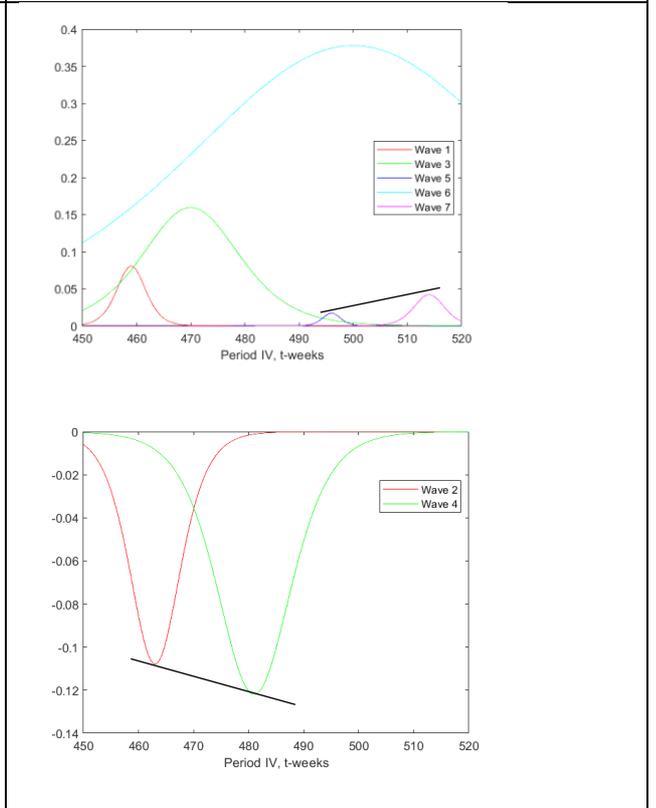
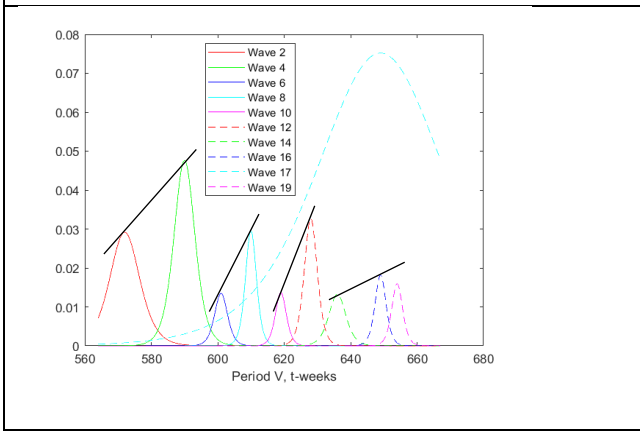
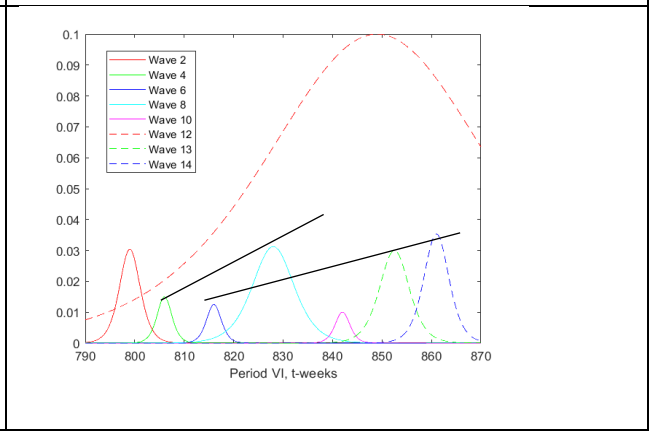

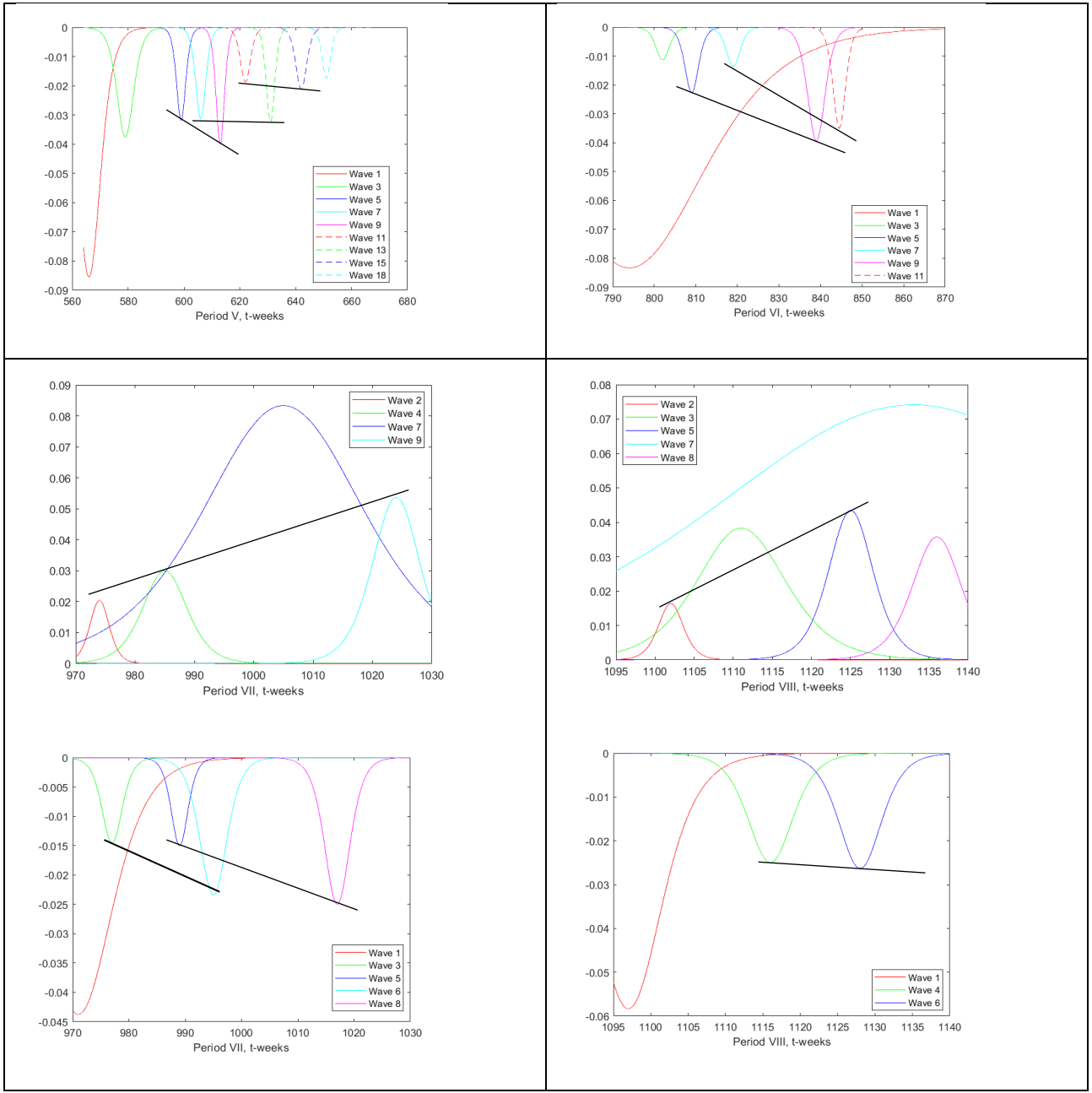

**Figure 7:** Waves in periods: I-VIII

CWT allows revealing the hidden structure of the trend. All periods can be decomposed into positive and negative parts, including one large (first-order) carrier wave and several smaller (second-order) waves indicating oscillations around the large wave. The second-order waves can be organized into one or more sequences. The amplitudes of the waves in each sequence grow linearly according to the relation $A_i - A_j = const * (T_i - T_j)$. In period I, two positive sequences and one negative sequence can be distinguished (marked with solid straight lines). The positive sequences can be associated with the positive part of the redundancy (Eq. 5), attributed to stabilization around the historical trajectory, and the negative part is related to evolutionary self-organization. The presence of two positive sequences indicates the existence of two separate, simultaneously existing regimes. Period II, describing a rapidly growing trend, is indicative, as it demonstrates one positive and one negative multi-wave sequence. Periods IV and VIII, which refer to slowing trends, are less representative in this regard. Only one positive and one negative sequence can be identified. Period V includes a highly volatile structure, which reflects multiple regimes of stabilization and self-organization. This also applies, to a lesser extent, to Period VI. Period VII marks a market top where stabilization processes form a second-order peak in a market decline. Period III is unique in that it contains a sequence of positive first-order waves, two sequences of positive second-order waves, and no sequences of negative waves. The point of the analysis is that we can 1) decompose a market trend into a nonlinear (first-order wave) and a "linearized" part (a sequence of linearly changing second-order waves); 2) interpret the resulting decomposition in terms of stabilizing and self-organizing processes; 3) use the "linearization" as a basis for more accurate forecasting of trend evolution.

## VII. Discussion

The results show that the model can accurately describe patterns in price charts, such as trends, and can provide insight into whether a trend will continue or stop. The nonlinear differential equation is considered as a basic method for describing price dynamics. Differential equations have previously been used by Caginalp and Balenovich (2003) to provide a basis for technical analysis. They showed that some technical analysis models can be modelled using differential equations when interactions between two or more groups of investors with different valuations and/or different motivational characteristics are assumed. However, no reasons were presented for what drives investor sentiment.

The presented model can be used to forecast a trend reversal. It is impossible to say for sure at the very beginning that a trend-like price movement is starting to reverse. But when the trend reverses and the wave structure is confirmed, certain price levels can be expected to be reached at certain time points. If the level is reached by the planned point, a (temporary) trend reversal can be expected. However, the presented model should not be considered a universal tool for forecasting financial time series in any periods of trend and price consolidation, given that Eq. 7 is derived under certain conditions that may not always be met.

The informational approach adds to a growing literature that examines the role of social norms, moral attitudes, religions, and ideologies in the imperfect rationality of market participants. Recently, there has been a growing understanding that investor thoughts and behaviour are influenced by accepted cultural traits, which can be viewed as an evolutionary system. This forms the basis of a new paradigm for understanding financial markets – social finance (e.g., Akçay and Hirshleifer, 2021). There are significant similarities between the social finance and informational approaches. Table 4 summarizes some of the main characteristics of the two approaches.

| Social finance | Information approach |
|---|---|
| Adopted cultural traits, including information signals, beliefs, strategies, and folk economic models are the drivers of investors' decisions | Sets of communication codes are the drivers of information processing and investors' decisions |
| Social transmission determine the evolution | Sets of communication codes shape each other |
| and mutation of financial traits | in the process of communication |
| Cultural traits are subject to different biases in judgments and decisions. | Different sets of communication codes provide different meanings to the same informational content |
| Cultural traits are not immediately measurable | Meaning can be measured on the base of the extension of Shannon's mathematical theory of communications |
| Social finance encompasses agent-based modeling of transmission biases | Information approach encompasses modelling the effect of meaning generation with help of non-linear evolutionary equation |
| Regards shifts in investors' sentiment as an endogenous outcome of microevolutionary cultural processes | Regards shifts in investors' sentiment as an endogenous outcome of interaction of communication codes sets, formed in the evolutionary process |
| Investors adopt and modify their financial traits | Agents adopt and modify their sets of communication codes |
| Considers a wider set of applications and range of time scales | Considers a wider set of applications and range of time scales |

| Cultural evolution operates at multiple time scales. The evolutionary dynamics of financial traits play out at multiple time scales | Communication code sets operate at multiple time scales. The evolutionary dynamics of meaning generation play out at multiple time scales |
|---|---|
| Is focused financial sphere | Can be applied a wider range of applications |

**Table 4.** Social finance and Information approach

The information approach complements the study of how social interaction shapes the thinking and behavior of market participants. Both social finance and the information approach consider the features of agent groups as the main cause of social transmission biases on different time scales. These features develop in the process of communication. There are also differences between the two approaches: 1) social finance explicitly focuses on how social interaction shapes the thinking and behaviour of investors while the information approach goes beyond finance and can be applied to a wider area of inter-social communications, such as innovation studies, the spread of infectious diseases, rumors propagation, etc. (Ivanova, 2024; Ivanova & Rzadkowski, 2024); 2) cultural traits are not directly evaluated, but the meaning can be measured as redundancy; 3) the information approach and social finance use different mathematical apparatus. The implementation of the nonlinear evolutionary equation technique allows for an adequate description of some of the observed price patterns and additionally provides the basis for technical analysis.

## VIII. Conclusion

This paper presents the first study of market price dynamics that examines the role of information processing and meaning generation in social systems. The study builds on Loet Leydesdorff's seminal work on the dynamics of expectations and meaning in interpersonal communication (Leydesdorff & Franse, 2009; Leydesdorff & Ivanova, 2014; Leydesdorff, Petersen & Ivanova, 2017). It also incorporates the conceptual framework of the triple helix model of university-industry-government relationships (Etzkowitz & Leydesdorff, 1995, 1998) and its mathematical formulation (Ivanova & Leydesdorff, 2014a, 2014b). The main contribution of the paper is that it represents a step forward in the development of the theory of meaning, which has potentially much broader applications than financial research. This paper can also be considered an example of applying the theory of meaning, which has long been considered an abstract discipline, to the specific area of financial markets.

Another contribution is the introduction of a quantitative model for studying the dynamics of price movements of market assets and forecasting future price movements based on a nonlinear evolutionary equation.

Meaning in social communication is processed through specific sets of communication codes that span horizons of meaning, acting as mechanisms of selection and coordination. It is based on expectations and emerges from events against the arrow of time. Expectations can be measured as redundancy (i.e. additional possibilities) using Shannon's entropic information theory. Conceptually, much can be gained by combining and testing these ideas with new methods and approaches. Perhaps even more important is the idea that meaning can be measured quantitatively. This is a very interesting and promising area of research that has great potential to expand our understanding of the dynamics of social systems and improve our prediction capabilities regarding events that have not yet occurred. It may prove useful when applied to various fields related to information exchange in social systems in general and to behavioral economics and financial markets in particular. The results of the article may also be useful for policy makers and market practitioners in their daily activities.

A subject for future research is to apply this approach to some other problems and datasets with different numbers of features.

**Appendix A**

Shannon informational entropy is defined as:
$$H = -\sum_{i=1}^{S} p_i \log p_i \tag{A1}$$

Dubois showed (2019) that taking into account temporal cyclic systems:
$$H = H(t) = -\sum_{i=1}^{S} p_i(t) \log p_i(t) \tag{A2}$$

with normalization conditions:
$$\frac{1}{T}\int_0^T \sum_{i=1}^{S} p_i(t) dt = 1, \quad H_0 = \frac{1}{T}\int_0^T H(t) dt \tag{A3}$$

in case $S = 2$ one obtains harmonic oscillator equation:
$$\begin{cases} \frac{dp_1}{dt} = -\frac{F}{p_{1,0}}(p_1 - p_{1,0}) \\ \frac{dp_2}{dt} = \frac{F}{p_{2,0}}(p_2 - p_{2,0}) \end{cases} \tag{A4}$$

where $p_{i0} = \frac{1}{T}\int_0^T p_i(t) dt$ and $F$ is any function of $p_i, t$. Following Dubois one can define the state of reference:
$$I_0 = -\sum_{i=1}^{S} p_{i,0} \log p_{i,0} \tag{A5}$$

and develop informational entropy $H$ in Taylor's series around the reference state:

$$H = I_0 - \sum_{i=1}^{S}[(\log p_{i,0} + 1)(p_i - p_{i,0}) + \frac{(p_i - p_{i,0})^2}{2 p_{i,0}} + \cdots O((p_i - p_{i,0})^3)] \tag{A6}$$

After substituting the Equation A5 into Equation A6 and neglecting the terms beyond the second degree one obtains:

$$H = -\sum_{i=1}^{S}[p_i \log p_{i,0} + (p_i - p_{i,0})] + D^* \tag{A7}$$

where:
$$D^* = \sum_{i=1}^{S}\left(\frac{(p_i - p_{i,0})^2}{2 p_{i,0}}\right) \tag{A8}$$

The condition for non-asymptotic stability of cyclic system is:

$$\frac{dD^*}{dt} = 0 \tag{A9}$$

In case $S = 2N$ one of possible solution of Equation A7 is:

$$\begin{cases} \frac{dp_{j-1}}{dt} = -\frac{\gamma}{p_{j,0}}(p_j - p_{j,0}) \\ \frac{dp_j}{dt} = \frac{\gamma}{p_{j-1,0}}(p_{j-1} - p_{j-1,0}) \end{cases} \tag{A10}$$

$j=2, 4, \ldots 2N$. Upon differentiating the system (A10) by time we obtain:

$$\begin{cases} \frac{d^2 p_j}{dt^2} = \frac{\gamma^2}{p_{j,0} p_{j-1,0}}(p_j - p_{j,0}) \\ \frac{d^2 p_{j-1}}{dt^2} = \frac{\gamma^2}{p_{j,0} p_{j-1,0}}(p_{j-1} - p_{j-1,0}) \end{cases} \tag{A11}$$

The function $D^*$ corresponds to the non-linear residue in (A6) which is a truncated version of (A5). Using non-truncated equations (A5) we obtain:

$$\begin{cases} \frac{d^2 p_j}{dt^2} = \frac{\gamma^2}{p_{j,0} p_{j-1,0}}(p_j - p_{j,0}) + C_j \\ \frac{d^2 p_{j-1}}{dt^2} = \frac{\gamma^2}{p_{j,0} p_{j-1,0}}(p_{j-1} - p_{j-1,0}) + C_{j-1} \end{cases} \tag{A12}$$

where $C_j = O(p_j - p_{j,0})^3_{tt}$. When $p_j$ are smaller than $p_{j-1}$, in order to keep the same order of magnitude one can drop the terms beyond the second degree for the variable $p_j$ and the terms beyond the third degree for the variable $p_{j-1}$. In a similar manner this leads to the function $D^{**}$ defined by analogy with $D^*$:

$$D^{**} = \sum_j^S \frac{(p_{j-1} - p_{j-1,0})^2}{2 p_{j-1,0}} + \frac{(p_j - p_{j,0})^2}{2 p_{j,0}} - \frac{(p_j - p_{j,0})^3}{6 p_{j,0}^2} \tag{A13}$$

Differentiating $D^{**}$ by time and equating to zero $\frac{dD^{**}}{dt} = 0$ we obtain a system:

$$\begin{cases} \frac{dp_{j-1}}{dt} = \frac{\gamma}{p_{j,0}}(p_j - p_{j,0}) - \frac{\gamma}{2p_{j,0}^2}(p_j - p_{j,0})^2 \\ \frac{dp_j}{dt} = -\frac{\gamma}{p_{j-1,0}}(p_{j-1} - p_{j-1,0}) \end{cases} \quad (A14)$$

which yields an equation for the non-harmonic oscillator:

$$\frac{d^2 p_j}{dt^2} = -\frac{\gamma^2}{p_{j,0} p_{j-1,0}}(p_j - p_{j,0}) + \frac{\gamma^2}{p_{j,0}^2 p_{j-1,0}}(p_j - p_{j,0})^2 + C'_j \quad (A15)$$

**Appendix B**

Redundancy (Equation 1) can be considered a result of a balance between two dynamics - evolutionary self-organization and historical organization (Leydesdorff, 2010). In other words it is a balance between recursion on a previous state on the historical axis as opposed to the meaning provided to the events from the perspective of hindsight (Dubois, 1998). Redundancy dynamics drives corresponding probabilities dynamics with recursive and incursive perspectives. Provided that probabilities oscillate in non-harmonic mode (Equation A15) one can write:

$$\frac{d^2 p_j}{dt^2} = -\frac{\gamma^2}{p_{j,0} p_{j-1,0}}(p_j^- - p_{j,0}^-) + \frac{\gamma^2}{p_{j,0}^2 p_{j-1,0}}(p_j^- - p_{j0}^-)^2 + \frac{\gamma^2}{p_{j,0} p_{j-1,0}}(p_j^+ - p_{j,0}^+) - \frac{\gamma^2}{p_{j,0}^2 p_{j-1,0}}(p_j^+ - p_{j,0}^+)^2 + C_j' + C_j'' \tag{B1}$$

here $p_j^-$ and $p_j^+$ are defined with respect to past and future states. Then using the trapezoidal rule we can write Equation A3 as:

$$p_{j,0}^- = \frac{1}{2}(p_j^- + p_j); p_{j,0}^+ = \frac{1}{2}(p_j^+ + p_j); \text{ so that } p_j^- - p_{j,0}^- = \frac{1}{2}(p_j - p_j^-); p_j^+ - p_{j,0}^+ = \frac{1}{2}(p_j^+ - p_j)$$

Developing $p_j^+$ and $p_j^-$ in Taylor's series in the state space:[9]

$$\begin{aligned} p_j^+ &= p_j + p_j' h + \frac{1}{2} p_j'' h^2 + \frac{1}{6} p_j''' h^3 + \frac{1}{24} p_j'''' h^4 + \cdots \\ p_j^- &= p_j - p_j' h + \frac{1}{2} p_j'' h^2 - \frac{1}{6} p_j''' h^3 + \frac{1}{24} p_j'''' h^4 + \cdots \end{aligned} \tag{B2}$$

and keeping the terms up to the $h^4$ order of magnitude one obtains (Fermi, Pasta, Ulam, 1955):

---

[9] The state space is presented by x axis

$$\frac{1}{k}p_{j_{tt}} = p_j'' h^2 + 2\alpha p_j' p_j'' h^3 + \frac{1}{12} p_j'''' h^4 + O(h^5) \tag{B3}$$

$$k = \frac{\gamma^2}{p_{j,0} p_{j-1,0}}; \ \alpha = \frac{1}{p_{j-1,0}}; \ C_1 = \frac{1}{k}(C_j' + C_j'')$$

Setting further: $w = \sqrt{k}$; $t' = wt$; $y = x/h$; $\varepsilon = 2\alpha$ one can rewrite Equation B3 in the form:

$$-p_{j_{tt}} + p_{j_{yy}} + \varepsilon p_{j_y} p_{j_{yy}} + \frac{1}{12} p_{j_{yyyy}} + C_1 = 0 \tag{B4}$$

Going to the moving coordinate system: $X = y - t'$, rescaling time variable $T = \frac{\varepsilon}{2}\tau$, and keeping terms up to the first order in $\varepsilon$ one brings Equation B3 to the form:

$$\varepsilon \Sigma_{XT} + \varepsilon \Sigma_X \Sigma_{XX} + \frac{1}{12} \Sigma_{XXXX} + C_1 = 0 \tag{B5}$$

Here $p_j = \Sigma(X, T)$. Defining further: $p = \Sigma_X$ and $\delta = \frac{1}{12\varepsilon}$ we obtain a non-linear evolutionary equation:

$$p_T + p p_X + \delta p_{XXX} + C_1 = 0 \tag{B6}$$

which corresponds to Korteweg de Vries (KdV) equation (Gibbon, 1985):

$$u_T + u u_X + \delta u_{XXX} = 0 \tag{B7}$$

with additional term $C_1$. By substitution: $\rightarrow 6P\ u \rightarrow 6U; T \rightarrow \sqrt{\delta}\tau;\ X \rightarrow \sqrt{\delta}\chi$ Equations B6 and B7 are reduced to the form:

$$P_\tau + 6PP_\chi + P_{\chi\chi\chi} + C_1 = 0 \tag{B8}$$

$$U_\tau + 6UU_\chi + U_{\chi\chi\chi} = 0 \tag{B9}$$

Equation B9 possesses soliton solutions:

$$U(\chi, \tau) = \frac{\kappa^2}{2} \text{sech}^2\left[\frac{k}{2}(\chi - \kappa^2 \tau)\right] \tag{B10}$$

the soliton solution of Equation B6 is:

$$P(X,T) = \frac{\kappa^2}{12}\operatorname{sech}^2\left[\frac{\kappa}{2\sqrt{\delta}}\left(X - \kappa^2 T + \frac{C_1}{2\sqrt{\delta}}T^2\right)\right] - C_1 T \qquad (B11)$$

It follows that for the train of solitons with amplitudes: $A_i$, $A_j$ and $A_k$ and times $T_i, T_j, T_k$ the following condition holds: $\frac{A_i - A_j}{T_i - T_j} = \frac{A_j - A_k}{T_j - T_k}$.

Equation (B8) after transition to a moving frame: $t = X - T$ and integration is written as follows:

$$P_{tt} + 3P^2 - P + C_1 t = 0 \qquad (B.12)$$

here $P$ stands for probability density.

We can further obtain the equation which governs redundancy evolution (Ivanova and Rzadkowski, 2024). By setting $P = e^q$ ($q < 0$) redundancy $R = P \ln P$ (we account that redundancy can be either positive or negative and not define a sign) can be presented in the form: $R = qe^q$. This expression can be inversed: $q = W(R)$, where $W(R)$ is the Lambert function (e.g. Lehtonen, 2016). Accordingly:

$$P = e^W = \frac{R}{W} \qquad (B.13)$$

Differentiation of (B.13) with respect to $t$ and accounting that:

$$W_t = \frac{W}{R(1+W)}R_t = \frac{R_t}{R + e^W}$$

$$W_{tt} = \frac{R_{tt}(R + e^W) - R_t^2 - R_t W_t e^W}{(R + e^W)^2}$$

yields:

$$P_{tt} = e^W W_t^2 + e^W W_{tt} = \frac{WRR_{tt}\left(1+\frac{1}{W}\right) - W\left(\frac{1}{1+W}\right)R_t^2}{R(1+W)^2} \qquad (B14)$$

Substituting (B14) into (B.12) we get:

$$\frac{W^3 RR_{tt}\left(\frac{1+W}{W}\right) - W^3 R_t^2 + 3R^3(1+W)^2 - R^3(1+W)^2}{W^2 R(1+W)^2} + C_1 t = 0 \qquad (B.15)$$

in a linear approximation for small $R$ values $W(R) \sim R$ (B.15) takes the form:

$$\frac{R_{tt} - \left(\frac{R_t}{1+R}\right)^2 + 2(1+R)}{(1+R)} + C_1 t = 0 \qquad (B.16)$$

Accounting that $(lnR)_t = \frac{R_t}{1+R}$, expanding the logarithm $R$ in a Tailor series, taking the derivative, squaring, and in a weakly non-linear assumption preserving terms up to the second order of smallness, we obtain:

$$(lnR)_t^2 \sim 1 - 2R + R^2 + \cdots \qquad (B.17)$$

Substituting (B.17) into (B.16), we obtain:

$$R_{tt} - R^2 + 4R + 1 + C_1 t = 0 \qquad (B.18)$$

Eq. (B.18) can be derived from the equation:

$$4R_T - 2RR_X + R_{XXX} + C_1 = 0 \qquad (B.19)$$

upon differentiation and transition to a moving frame $t = X + T$.

**Appendix C**

In supposition of existence a periodic solution (Gibbon, 1985):

$$U(x,t) = f(x - vt) \qquad (C1)$$

substitution of expression C1 into Equation B7 yields:

$$-vf' + ff' + f''' = \frac{1}{2}a \qquad (C2)$$

here $a$ is a constant of integration. Multiplying Equation C2 by $2f'$ and integrating it one obtains an equation which has periodic solutions in the form of elliptic functions:

$$f'^2 = -\frac{1}{3}f^3 + vf^2 + af + b \qquad (C3)$$

Corresponding solution for Equation B7 then takes the form:

$$P(x,t) = f(x - vt + Ct^2/2) - Ct \qquad (C4)$$

## Appendix D

For the nth derivative $x^n(t)$ of the basic logistic function $x(t) = \frac{1}{1+e^{-t}}$ it holds

$$\int_{-\infty}^{\infty} \left(x^{(n)}(t)\right)^2 dt = (-1)^{n-1} B_{2n} = |B_{2n}| \tag{D1}$$

where $B_{2n}$ is the $(2n)$th Bernoulli number.

*Proof.*

Examining the soliton solutions of the Korteweg-de Vries equation, Grosset and Veselov (2005) obtained an interesting relationship between these solutions and the Bernoulli numbers

$$\int_{-\infty}^{\infty} \left(\frac{d^{n-1}}{dt^{n-1}} \frac{1}{\cosh^2 t}\right)^2 dt = (-1)^{n-1} 2^{2n+1} B_{2n} \tag{D2}$$

Other proofs of the Grosset-Veselov formula (D2) can be found in (Boyadzhiev, 2007; Rzadkowski, 2010). The formula (D1) follows from the formula (D2) because (we put $\tau = 2t$ at the end):

$$\int_{-\infty}^{\infty} \left(\frac{d^{n-1}}{dt^{n-1}} \frac{1}{\cosh^2 t}\right)^2 dt = \int_{-\infty}^{\infty} \left(\frac{d^{n-1}}{dt^{n-1}} \frac{4e^{-2t}}{(1+e^{-2t})^2}\right)^2 dt = 4 \int_{-\infty}^{\infty} \left(\frac{d^n}{dt^n} \frac{1}{1+e^{-2t}}\right)^2 dt =$$

$$4(2^n)^2 \int_{-\infty}^{\infty} \left(x^{(n)}(2t)\right)^2 dt = 2(2^n)^2 \int_{-\infty}^{\infty} \left(x^{(n)}(\tau)\right)^2 d\tau \tag{D3}$$

Comparing (D3) with (D1) we get (D2).

Bernoulli number $B_n$ vanishes for all odd numbers $n \geq 3$. The first few non-zero Bernoulli numbers are as follows $B_0 = 1$, $B_1 = -\frac{1}{2}$, $B_2 = \frac{1}{6}$, $B_4 = -\frac{1}{30}$, $B_6 = \frac{1}{42}$, $B_8 = -\frac{1}{30}$, $B_{10} = \frac{5}{66}$, $B_{12} = -\frac{691}{2730}$ (see Duren (2012), Ch. 11).

Since $B_4 = -1/30$, than from equation (18) it follows that the normalized mother wavelet $\psi_2(t)$ is of the form

$$\psi_2(t) = \sqrt{30}\, x''(t) = \frac{\sqrt{30}(e^{-2t}-e^{-t})}{(1+e^{-t})^3} \tag{D4}$$

**Appendix E**

Let a function $f(t)$ be of the form:

$$f(t) = c + dt + \frac{x_{sat}}{1+\exp\left(-\frac{t-b}{a}\right)}$$

*The continuous wavelet transform (16) of the function $f''(t)$, by using the logistic second-order wavelets $\psi_2^{\alpha,\beta}$ (19)*

$$(T^{wav}f'')(\alpha,\beta) = \langle f'', \psi_2^{\alpha,\beta}\rangle = \int_{-\infty}^{\infty} f''(t)\, \psi_2^{\alpha,\beta}(t)\, dt$$

*takes maximum (for $x_{sat} > 0$) or minimum (for $x_{sat} < 0$) value when $\alpha = a$ and $\beta = b$.*

*Proof.*

Using definition (19) of the mother second-order logistic wavelet it is easy to check that

$$f''(t) = \frac{x_{sat}}{\sqrt{30}\, a^{3/2}}\, \psi_2^{a,b}(t)$$

Assume that $x_{sat} > 0$. By the Cauchy-Schwartz inequality

$$|(T^{wav}f'')(\alpha,\beta)| = |\langle f'', \psi_2^{\alpha,\beta}\rangle| \le \|f''\|\|\psi_2^{\alpha,\beta}\| = \frac{x_{sat}}{\sqrt{30}\,a^{3/2}}\|\psi_2^{a,b}\|\|\psi_2^{\alpha,\beta}\| = \frac{x_{sat}}{\sqrt{30}\,a^{3/2}}$$

The maximum is reached for $\alpha = a$, $\beta = b$, because:

$$(T^{wav}f'')(a,b) = \langle f'', \psi_2^{a,b} \rangle = \frac{x_{sat}}{\sqrt{30}a^{3/2}} \langle \psi_2^{a,b}, \psi_2^{a,b} \rangle = \frac{x_{sat}}{\sqrt{30}a^{3/2}} \tag{E1}$$

Similarly we consider the case $x_{sat} < 0$.